\documentclass[twocolumn,prd,nofootinbib,preprintnumbers]{revtex4}

\pdfoutput=1
\usepackage{graphicx}
\usepackage{amsmath,bm}
\usepackage[T1]{fontenc}
\usepackage{rotating}
\usepackage{float}
\usepackage{aas_macros}
\floatstyle{plaintop}

\usepackage{color}
\definecolor{naviBlue}{RGB}{0,0,128}
\usepackage[colorlinks  = true,
            linkcolor   = naviBlue,
            urlcolor    = naviBlue,
            citecolor   = naviBlue,
            anchorcolor = naviBlue]{hyperref}
\newcommand{\secref}[1]{\hyperref[sec::#1]{SECTION~\ref*{sec::#1}}}
\newcommand{\subsecref}[1]{\hyperref[subsec::#1]{SECTION.~\ref*{subsec::#1}}}
\newcommand{\figref}[1]{\hyperref[fig::#1]{FIG.$\,$\ref*{fig::#1}}}
\newcommand{\tabref}[1]{\hyperref[tab::#1]{TABLE$\,$\ref*{tab::#1}}}
\newcommand{\eqnref}[1]{\hyperref[eqn::#1]{Eq.$\,$(\ref*{eqn::#1})}}

\newcommand{\diff}{\mathrm{d}}

\newcommand{\eq}{\mathrm{eq}}

\def\beq{\begin{equation}}
\def\eeq{\end{equation}}
\definecolor{darkgreen}{RGB}{0,170,0}
\definecolor{darkgray}{RGB}{110,110,108}
\newcommand{\bea}{\begin{eqnarray}}
\newcommand{\eea}{\end{eqnarray}}
\newcommand{\be}{\begin{equation}}
\newcommand{\ee}{\end{equation}}
\newcommand{\nn}{\nonumber}

\usepackage{array}
\newcolumntype{C}{>{$}c<{$}} 	%use math mode in table column by default

\definecolor{purple}{RGB}{160,0,160}
\definecolor{plotpink}{RGB}{205,0,180}
\definecolor{plotcyan}{RGB}{0,215,215}
\definecolor{darkcyan}{RGB}{17,155,155}

\definecolor{plotblue}{RGB}{0,0,235}
\definecolor{plotorange}{RGB}{245,140,0}
\definecolor{darkorange}{RGB}{210,100,0}
\definecolor{plotgreen}{RGB}{30,130,0}
\definecolor{plotred}{RGB}{240,0,0}
\definecolor{darkred}{RGB}{180,0,0}
\definecolor{darkergreen}{RGB}{0,130,3}
\definecolor{darkgreen}{RGB}{0,160,0}
\definecolor{lblue}{RGB}{100,130,205}
\definecolor{ourbrown}{RGB}{151,105,56}
\definecolor{darkblue}{RGB}{10,10,145}
\definecolor{gray}{RGB}{90,90,90}
\definecolor{graycyan}{RGB}{90,113,113}
\definecolor{dgraycyan}{RGB}{90,113,113}
\definecolor{darkgraycyan}{RGB}{66,84,84}
\definecolor{darkH}{RGB}{11,65,188}

\begin{document}

\title{
Interplay of super-WIMP and freeze-in production of dark matter
}

\author{Mathias Garny}
\affiliation{Physik Department T31, Technische Universit\"at M\"unchen,
James-Franck-Stra\ss e 1,
D-85748 Garching, Germany}
\author{Jan Heisig}
\affiliation{Institute for Theoretical Particle Physics and Cosmology, RWTH Aachen University, Sommerfeldstra\ss e 16, D-52056 Aachen, Germany}
\preprint{TUM-HEP 1166/18}
\preprint{TTK-18-39}

\begin{abstract}
Non-thermalized dark matter is a cosmologically valid alternative to the paradigm of weakly interacting massive particles. For dark matter belonging to a $Z_2$-odd sector that contains in addition a thermalized mediator particle, dark matter production proceeds in general via both the freeze-in and super-WIMP mechanism. 
We highlight their interplay and emphasize the connection to long-lived particles at colliders. For the explicit example of a colored $t$-channel mediator model we map out the entire accessible parameter space, cornered by bounds from the LHC, big bang nucleosynthesis and Lyman-$\alpha$ forest observations, respectively. We discuss prospects for the HL- and HE-LHC\@. 
\end{abstract}

\maketitle

%===================================================================
\section{Introduction}\label{sec:intro}
%===================================================================

The evidence for dark matter (DM) in our Universe provides a strong motivation for extending the standard model (SM) of particle physics by a dark sector containing a thermally or non-thermally produced relic. While the hypothesis of a thermalized and frozen-out DM candidate -- such as a weakly interacting massive particle (WIMP) -- is an attractive and thus widely studied possibility, it is by far not the only viable explanation. In particular, in view of many null-results from WIMP searches, an exploration of alternative scenarios is vital to exploit the current experimental capabilities and identify the nature of DM\@. 

One such scenario is feebly interacting DM that never reaches thermal equilibrium with the SM throughout the cosmological history. In this case DM production may proceed via  occasional scatterings or decays of particles in the thermal bath~\cite{Bolz:2000fu,Pradler:2006hh,McDonald:2001vt,Covi:2002vw,Asaka:2005cn}, so-called freeze-in~\cite{Hall:2009bx}. Another possibility is the out-of-equilibirum decay of a thermally decoupled mother particle, i.e.~through the super-WIMP mechanism~\cite{Covi:1999ty,Feng:2003uy}. The latter is realized in models where the mother particle belongs to a $Z_2$-odd dark sector, that forbids its decay into SM particles, while it may have sizeable couplings to the SM\@. In this case, the mother particle freezes out similarly to a WIMP while DM is produced through its decay, that typically becomes efficient much later in cosmic history. In addition, in general, a contribution to DM production from freeze-in is also present within this setup, as long as the mediator decay is possible~\cite{Hall:2009bx}. 

In this article we highlight the phenomenological implications of the interplay of super-WIMP and freeze-in production of DM and provide up-to-date experimental constraints and prospects. We consider a $Z_2$-odd 
dark sector comprising a feebly interacting DM particle and a mediator that transforms non-trivially under the SM gauge groups, such that its gauge interactions drive it towards thermal equilibrium in the early Universe. In contrast, the feeble DM interactions prevent it from thermalizing. For concreteness, we focus on a Majorana fermion DM candidate and a colored $t$-channel mediator, mapping out the entire accessible parameter space.

The DM density constraint imposes a fairly general relation between the involved masses and the lifetime of the mediator~\cite{Hall:2009bx}. In a wide range of the cosmologically allowed parameter space the mediator has macroscopic decay length allowing for experimental tests for long-lived particles at colliders as well as effects on big bang nucleosynthesis (BBN) through late decaying mediators. They constrain the parameter space towards small mediator masses and small mass splittings, respectively. We discuss current constraints from searches for detector-stable $R$-hadrons as well as future projections for stable and metastable mediators at the HL- and HE-LHC. For large mass splittings and a significant super-WIMP contribution to DM production, large deviations of the 
DM momentum distribution from the thermal one can arise. This leads to a large free-streaming length, suppressing the amplitude of the matter power spectrum on small scales, which can be probed via Lyman-$\alpha$ forest observations~\cite{Palanque-Delabrouille:2013gaa,2016A&A...594A..91L,Viel:2013apy}. This constrains the parameter space towards large mass splittings. The same observation constrains the parameter space towards very small DM masses (few keV) where freeze-in dominates. The parameter space is hence cornered from all sides.

The remainder of this work is structured as follows. We first introduce the model under consideration in Sec.~\ref{sec:model} and refer to 
possible embeddings and variations. In Sec.~\ref{sec:fifo} we detail the computation of the DM density and provide some model-independent phenomenological considerations.
Finally, Sec.~\ref{sec:constr} provides results for the cosmologically viable parameter space, experimental constraints and future projections. We conclude in Sec.~\ref{sec:summary}.

%===================================================================
\section{The model}\label{sec:model}
%===================================================================

As a simple example of a $Z_2$-odd new physics sector we consider a top-philic, colored scalar $t$-channel mediator 
$\tilde t$ and a Majorana DM fermion $\chi$ interacting with the SM through the Lagrangian
\begin{equation}
    \mathcal{L}_\text{int} = |D_\mu \tilde t|^2 + \lambda_\chi \tilde t\, \bar{t}\,\frac{1-\gamma_5}{2}\chi +\text{h.c.}\,,
    \label{eq:stopmodel}
\end{equation}
where $D_\mu$ is the covariant derivative, $t$ is the top quark Dirac field and $\lambda_\chi$ is the new
physics coupling. The $\tilde t$ particle is a $SU(2)_L$ singlet and has hypercharge identical to $t_R$, similar
to a right-handed squark field in supersymmetry.
The model introduces the three parameters $m_\chi$, $m_{\tilde t}$ and $\lambda_\chi$.

In this work we focus on the regime of sufficiently small couplings $\lambda_\chi$, such that the $\chi$ particle
never reaches thermal equilibrium with the SM bath. This means that neither $\tilde t - \chi$ conversions,
such as (inverse) decays $\tilde t\leftrightarrow t\chi$, nor annihilations, such as $\chi\chi\leftrightarrow t\bar t$,
occur at rates comparable to the Hubble expansion rate throughout cosmic history.
Details on further possible interactions and its phenomenology in the case of thermalized DM
can be found in~\cite{Ibarra:2015nca,Garny:2018icg}.

The simplified model, and variants with different spin- and gauge-quantum numbers, can be part of
generic extensions of the SM of particle physics. For example, the model possesses a natural
embedding in supersymmetric models. In this case the $Z_2$-symmetry can be identified with $R$-parity,
and the mediator with the lightest superpartner of the SM particles (being the right-handed stop for the specific model
from above). The feebly interacting DM particle can be realized in the context of a hidden sector,
that features an unbroken hidden $U(1)$ gauge symmetry. After supersymmetry breaking, a small kinetic 
mixing with the SM $U(1)_Y$ hypercharge leads to a small bino-admixture of the hidden gaugino, providing a small coupling $\lambda_\chi$ of the
form introduced above~\cite{Ibarra:2008kn,DeSimone:2010tr}.

Arguably, also supersymmetric models featuring gravitino DM and a long-lived next-to lightest supersymmetric particle 
(NLSP) share similarities with the type of models studied here if $R$-parity is conserved, but also exhibit differences due to the
non-renormalizeable interactions \cite{Bolz:2000fu,Pradler:2006hh} (see \cite{Covi:2014fba} for a recent analysis of stop NLSP, and
references therein for other possibilities).

A variant of the model considered here, but without $Z_2$-symmetry, has been studied in~\cite{Arcadi:2013aba,Arcadi:2014tsa}.

%===================================================================
\section{Dark matter production}\label{sec:fifo} 
%===================================================================

For small enough values of the coupling $\lambda_\chi$ that connects DM to the SM,
the DM particle $\chi$ is never in equilibrium with the SM thermal bath.
In this case, any process throughout the cosmic history leading to the production of $\chi$ particles
contributes to an accumulated $\chi$ population. Immediately after the end of inflation, $\chi$
particles may be produced during the reheating process. In this work we assume that this process
leads to a negligible contribution to the abundance of $\chi$ particles, and adopt the common assumption that 
reheating produces a thermal bath of SM particles, with maximal temperature given by $T_R$.
Furthermore, we assume $T_R\gg m_{\tilde t}$, such that the mediator $\tilde t$ thermalizes due to its
gauge interactions.\footnote{For $T_R\lesssim m_{\tilde t}$ the relic density becomes dependent on $T_R$ and
the production via freeze-in may dominantly proceed via DM pair production whose rate is 
suppressed by heavy mediator propagators arising in the $t$-channel or in loops. Hence, significantly larger
couplings are expected to saturate the relic density constraint than found in this work.}
In this case, within the simple model considered here, there are two distinct sources
of $\chi$ particle production. First, the freeze-in mechanism that is most efficient for $T\sim m_{\tilde t}$,
and second, the super-WIMP mechanism, corresponding to the late decay of the frozen-out population of $\tilde t$.
In the following we discuss both sources in turn.

%------------------------------------------------------------------------------------------
\subsection{Freeze-in}\label{sec:fi}
%------------------------------------------------------------------------------------------

Freeze-in production relies on the occasional production of $\chi$ particles within a thermal
bath. For the model considered here, due to the $Z_2$-symmetry in the dark sector, production
processes have to involve $\tilde t$ in the initial or final state. Since the abundance
of $\tilde t$ becomes strongly suppressed for $T\ll m_{\tilde t}$, the relevant temperature
range for freeze-in is $T\gtrsim m_{\tilde t}$. At these temperatures, gauge interactions
keep $\tilde t$ close to thermal equilibrium, i.e.~we may assume $n_{\tilde t}\simeq n_{\tilde t}^{\text{eq}}=g_{\tilde t}\int\frac{d^3p}{(2\pi)^3}f_{\tilde t}$
where $f_{\tilde t}=(e^{E_{\tilde t}/T}-1)^{-1}$, $g_{\tilde t}=N_c$
and $E_{\tilde t}^2=m_{\tilde t}^2+p^2$.

We consider both the $1\to 2$ process $\tilde t\to\chi t$ as well as
all allowed $2\to 2$ processes $ab\to \chi c$, 
including $\tilde t \bar t\to \chi g$,  $\tilde t g\to \chi t$,  $g \bar t\to \chi \tilde t^*$.
The Boltzmann equation for the number density $n_\chi$ reads~\cite{Hall:2009bx}
\begin{equation}\label{eq:Boltzmann}
\dot n_\chi + 3n_\chi H = 2\left(C_{1\to 2}+C_{2\to 2}\right)\,.
\end{equation}
Here
\bea
  C_{1\to 2} &=& \int\frac{\diff^3 p_{\tilde t}}{(2\pi)^3}  \frac{ g_{\tilde t} f_{\tilde t} m_{\tilde t}}{E_{\tilde t}}\Gamma_{\tilde t \to \chi t} = n_{\tilde t}\left\langle \frac{m_{\tilde t}}{E_{\tilde t}} \right \rangle \Gamma_{\tilde t \to \chi t} \,, \nn\\
  C_{2\to 2} &=& \!\!\!\!\sum_{\text{processes}}\int\frac{\diff^3 p_a}{(2\pi)^3}\frac{\diff^3 p_b}{(2\pi)^3}g_af_ag_bf_b\sigma_{ab\to\chi c}v_{ab}\,,
\eea
where $v_{ab}=\sqrt{(p_a\cdot p_b)^2-m_a^2m_b^2}/(E_aE_b)$, $g_a$ are the internal degrees of freedom of species $a$,
and we neglected the loss term, as appropriate for freeze-in, as well as statistical factors $(1\pm f_i)$ (see below).  
The factor $2$ in \eqref{eq:Boltzmann} takes into account
charge conjugated processes, which contribute equally due to CP symmetry and the Majorana nature of $\chi$.
Since $n_\chi$ appears only on the left-hand side, the Boltzmann equation can be solved by direct integration
for the yield $Y_\chi=n_\chi/s$, where $s=\frac{\pi^2}{45}g_{*S}T^3$ is the entropy density. The final yield 
\be
Y_\chi^\text{fi}=Y_\chi^{1\to 2}+Y_\chi^{2\to 2}
\ee
can be split into contributions from $1\to2$ and $2\to2$ processes.
The former is, for example, given by 
\begin{equation}
\label{eq:fisol}
Y_\chi^{1\to 2}(x_0) \simeq \int_0^{x_0}\!\diff x\, \frac{2Y_{\tilde t}^\text{eq}(x)}{x H(x)}\langle \Gamma\rangle(x)\,,
\end{equation}
where $x=m_{\tilde t}/T$, we assumed $\diff g_{*S}/\diff x=0$ during freeze-in and introduced
\begin{equation}
\langle \Gamma \rangle (x)= \Gamma_{\tilde t \to \chi t} \left\langle \frac{m_{\tilde t}}{E_{\tilde t}} \right \rangle\!(x) =\Gamma_{\tilde t \to \chi t}\frac{K_1(x)}{K_2(x)}\,,
\end{equation}
where $K_i$ denote modified Bessel functions.
The integral saturates for $x_0\gtrsim 1$ due to Boltzmann suppression of $Y_{\tilde t}^\text{eq}$.
The contribution to the DM density from freeze-in is given by
$(\Omega h^2)^\text{fi}=m_\chi Y_\chi^\text{fi}s(T_0)h^2/\rho_\text{crit}$,
where $s(T_0)$ and  $\rho_\text{crit}$ denote the entropy- and critical energy density today, respectively.

We compute the freeze-in contribution with \textsc{micrOMEGAS}~5.0.4~\cite{Belanger:2018ccd}
which assumes the mediator to follow the equilibrium density. As discussed above
we expect this to be a good approximation for the setup considered here. 
We used the default approximate phase space integration as well as the optional full vegas integration routine,
that includes also quantum statistical factors, and found deviations below $5\%$.

For $m_{\tilde t} > m_t+m_\chi $, the two-body decay $\tilde t\to \chi t$ is kinematically allowed.
Since $2\to2$ processes are formally suppressed by two powers of a SM coupling constant,
one may expect them to give a subdominant contribution in that case.
Nevertheless, we find them to contribute at the same level as the decay.
This has several reasons. For processes such as
$\tilde t \bar t\to \chi g$,  $\tilde t g\to \chi t$,  $g \bar t\to \chi \tilde t^*$
the relevant SM coupling is $\alpha_s$, which is sizeable for most of the parameter
space. In addition, $2\to 2$ processes are favored kinematically over $1\to 2$
for $T\gtrsim m_{\tilde t}$. Lastly, there is a large number of possible $2\to 2$
processes that add up, while only a single $1\to 2$ channel exists.

When both $1\to 2$ and $2\to 2$ processes are kinematically allowed, unphysical divergences may occur related
to nearly on-shell propagators (the default routine within \textsc{micrOMEGAS}
excludes $2\to2$ processes in that case). We checked that no such effects occur at a 
sizeable level, except for  $\tilde t g\to\chi t$ and $\tilde t \gamma\to\chi t$. For these processes the cross-section $\sigma(s)$ 
becomes enhanced close to the threshold $\sqrt{s}\gtrsim m_{\tilde t}$.
These processes feature an $s$-channel contribution, involving a propagator $1/(s-m_{\tilde t}^2)$ that gives a
large contribution close to threshold. A similar effect occurs for $\tilde t Z\to\chi t$ when $m_{\tilde t}\gg m_Z$.
The enhancement can be understood as soft initial state radiation that contributes to the next-to leading
order correction to the decay $\tilde t\to \chi t$, in an expansion in the SM couplings. 
We expect it to be regulated for the $\chi$ production rate when consistently including all real and virtual corrections.
In addition, one may argue that for the processes above the enhancement is cut off when including a thermal mass
for the soft initial state particle. Here we do not attempt to provide a full next-to leading order result.
Instead, we implement a cut-off $\sqrt{s}_\text{min}=R(m_a+m_b)$, where $R>1$. We checked that as long as $R$ is close to unity
the final yield depends only very weakly on the precise choice. We used $R=1.2$ in our numerical results.
The total value of the final yield is affected at most at the $10\%$ level for $m_{\tilde t}\lesssim 10^5$\,GeV
when choosing $R=1.1$ instead.
In addition, we checked that when omitting the processes $\tilde t g\to\chi t$, $\tilde t \gamma\to\chi t$
and  $\tilde t Z\to\chi t$ in the abundance calculation, 
all results remain qualitatively unchanged (see Sec.\,\ref{sec:ps} for details).

%------------------------------------------------------------------------------------------
\subsection{Super-WIMP}
%------------------------------------------------------------------------------------------

The super-WIMP mechanism relies on the thermal freeze-out of the mediator, that
subsequently decays into the DM particle~\cite{Feng:2003uy}. Within the model considered here, freeze-out
of $\tilde t$ annihilation into SM particles yields an abundance $Y_{\tilde t}^\text{fo}$
at temperatures $T\ll m_{\tilde t}/25$, analogous to WIMP freeze-out, that can be converted
into the density parameter $(\Omega h^2)_{\tilde t}=m_{\tilde t}Y_{\tilde t}^\text{fo}s(T_0)h^2/\rho_\text{crit}$.
At a much later time $t\simeq 1/\Gamma_{\tilde t\to \chi t}$, the mediator
decays, which yields a contribution to the DM density given by
\be
\label{eq:sWcont}
(\Omega h^2)^\text{sW}_\chi = m_\chi/m_{\tilde t}\, (\Omega h^2)_{\tilde t}\,.
\ee
We compute the freeze-out abundance of the mediator 
in the absence of DM as a function of $m_{\tilde t}$
using \textsc{micrOMEGAS}~5.0.4~\cite{Belanger:2018ccd}.
We take into account Sommerfeld enhancement of the mediator
annihilation cross section as detailed in Appendix B of~\cite{Garny:2017rxs}.
The mediator density may also be affected by bound-state effects
\cite{Liew:2016hqo,Mitridate:2017izz,Biondini:2018pwp}, which we leave for future work.

%------------------------------------------------------------------------------------------
\subsection{Some model-independent considerations}\label{sec:moind}
%------------------------------------------------------------------------------------------

Before discussing the parameter space and phenomenology of the specific model
considered here in more detail, we highlight several properties that apply
more generally to models containing a $Z_2$-odd mediator with sizeable interactions with
the SM, along with a DM particle that is very weakly coupled.
These types of models are not constrained by WIMP searches for direct- and indirect detection.
However, the presence of the mediator leads to testable signatures. It is in thermal equilibrium 
in the Early Universe, and can potentially be produced in laboratory experiments.
The mediator needs to be heavier than DM, such that it
can decay into the stable DM state. Due to the weak coupling, the mediator decay rate is suppressed, leading
generically to long-lived particles with special implications for phenomenology.

In this case both freeze-in and super-WIMP contributions are present in general.
The former depends on the production rate, which in turn depends on the small DM
coupling, in our case $(\Omega h^2)^\text{fi}_\chi\propto \lambda_\chi^2$. Since, for small
enough $\lambda_\chi$, this coupling plays no role for the mediator freeze-out,
the super-WIMP contribution is independent of the DM coupling, 
i.e.~$(\Omega h^2)^\text{sW}_\chi\propto \lambda_\chi^0$.
Therefore, if we require that the total abundance matches the observed value,
$(\Omega h^2)^\text{fi}_\chi(\lambda_\chi)+(\Omega h^2)^\text{sW}_\chi=0.12$,
solutions can only exist for points in parameter space for which $(\Omega h^2)^\text{sW}_\chi\leq 0.12$.
In that case, the condition above can be used to determine the value of $\lambda_\chi$ to explain
the measured DM density. We therefore expect in general that within a large portion of the parameter
space freeze-in dominates  (``bulk''). The viable region in parameter space is then bounded by a hypersurface on which
the super-WIMP mechanism saturates the DM density constraint (``boundary''). We will see below that this expectation is
borne out in the model considered here (cf.~\cite{Heisig:2018teh}).

Provided the process that corresponds to mediator decay gives a sizeable contribution to the
freeze-in abundance (in the model considered here this is the case for $m_{\tilde t}>m_t+m_\chi$,
such that two-body decay is kinematically allowed), we can estimate the freeze-in
abundance by generalizing~\eqref{eq:fisol} in the form 
\begin{equation}
\label{eq:fisol2}
Y_\chi^{1\to 2} \simeq c\,g_\text{med}\frac{\Gamma_{\text{med}}}{m_{\text{med}}^2}
\int_0^{x_0}\!\diff x\, \frac{Y_{\text{med}}^\eq(x)}{x \widetilde H(x)g_\text{med}}\frac{K_1(x)}{K_2(x)}\,,
\end{equation}
where $\widetilde H =H/m_{\text{med}}^2$, and $c=1(2)$ for a neutral (charged) mediator. 
Within the model considered here, $\Gamma_\text{med}\to \Gamma_{\tilde t \to \chi t}$,
$m_\text{med}\to m_{\tilde t}$, $g_\text{med}\to g_{\tilde t}$, $c\to 2$.

As long as the temperature of mediator freeze-out is well above the electroweak scale
the number of relativistic degrees of freedom is approximately constant such that
$Y_{\text{med}}^\eq(x)$ and $\widetilde H(x)$ are functions of $x$ only, without reference to $m_{\text{med}}$.
Furthermore, the number of internal degrees of freedom of the mediator cancels out inside the integrand 
in Eq.~\eqref{eq:fisol2}. Hence, the integral in Eq.~\eqref{eq:fisol2} is a constant. 
Consequently, within the ``bulk'' region of parameter space (for which $(\Omega h^2)^\text{sW}_\chi\leq 0.12$),
\begin{equation}
\label{eq:fisol3}
\frac{\Omega h^2}{0.12} \simeq 8.5\times10^{24} \, c\,g_{\text{med}}  \Gamma_{\text{med}}\frac{m_\chi}{m_{\text{med}}^2}\,,
\end{equation}
where we negected the $2\to 2$ contribution in order to obtain the parametric estimate above (cf.~\cite{Hall:2009bx}).
For a given decay rate $\Gamma_{\text{med}}$, or equivalently mediator lifetime, this imposes
a correlation $m_{\text{med}}\propto m_\chi^{1/2}$ between the mediator and DM mass.
We expect this finding to be applicable to the general class of models discussed above, see e.g.~\cite{Bernal:2017kxu}.

In addition, within the ``bulk'' region of parameter space, for which freeze-in dominates,
Eq.\,\eqref{eq:fisol3} can be used to estimate the time $t_\text{dec}\simeq \Gamma_{\text{med}}^{-1}$ 
when the (sub-dominant) population of frozen-out mediator particles decays. 
In terms of temperature, and well above the electroweak scale, $T_\text{dec}\simeq 6\cdot 10^8\,\text{GeV} \times(\Gamma_{\text{med}}/\text{GeV})^{1/2}$
and hence $x_\text{dec}=m_{\text{med}}/T_\text{dec}$ is given by
\begin{equation}
x_\text{dec} \simeq 378 \,\left(c\,g_{\text{med}}\right)^{1/2} \left(\frac{m_\chi }{\text{MeV}}\right)^{1/2}\,.
\end{equation}
That is, there is a relation $m_\chi\propto x_\text{dec}^2$, which we again expect to apply
to the class of models discussed above. This also shows that the super-WIMP production via mediator decay is well
separated in time from the freeze-in regime $x\simeq {\cal O}(1)$ for $m_\chi \gg 0.1$\,keV.

Note that Eq.~\eqref{eq:fisol3} furthermore implies a model-independent statement about the region in the 
mediator-DM mass plane that provides long-lived particles at the LHC\@. For proper decay length in the range 
$[1\,\text{m}\;\!;\,1\,\text{mm}]$ we find
\begin{equation}
\label{eq:metastable}
m_{\text{med}}\simeq[1.3;\,40]\,\text{TeV}\,\left(c\,g_{\text{med}}\right)^{1/2}\left(\frac{m_\chi }{\text{MeV}}\right)^{1/2}\,,
\end{equation}
where the lower edge of the mass range corresponds to the upper edge of the decay length and vice versa
while smaller masses provide mostly detector-stable mediators. 
Note that in case of additional contributions to DM production the lifetime becomes larger, shifting
the respective mediator mass range to larger values.
Within freeze-in scenarios long-lived particle signatures at the LHC were studied in~\cite{Hall:2009bx,Co:2015pka,Hessler:2016kwm,Ghosh:2017vhe,DEramo:2017ecx,Brooijmans:2018xbu,Calibbi:2018fqf}.

%=======================
%    \                                           |
%      \                                         |
%        \                                       |
\begin{figure*}[t]
\centering
\setlength{\unitlength}{1\textwidth}
\begin{picture}(0.96,0.34)
\put(0.0,-0.012){\includegraphics[trim={75 100 80 140},clip,width=0.96\textwidth]{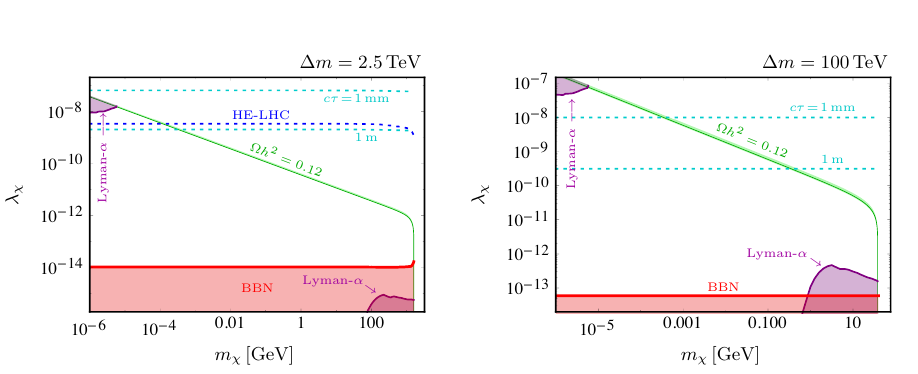}}
\end{picture}
\caption{
Dark matter coupling strength $\lambda_\chi$ providing $\Omega h^2 \simeq 0.12$ as a function of
$m_\chi$ (green solid curve) for $\Delta m = 2.5$\,TeV (left) and 100\,TeV (right). The green shaded band provides an estimate of the uncertainties due 
to the regularization of divergent diagrams: the upper boundary is obtained from leaving out the
respective diagrams while the lower boundary represents a looser cut (see text for details).
The purple and red shaded regions denote 95\% CL exclusions from Lyman-$\alpha$ and BBN bounds,
respectively. 
The cyan dotted curves show contours of constant mediator decay length, while the blue dotted curve shows the
projected sensitivity of the 27\,TeV HE-LHC\@.
}
\label{fig:paramscan1}
\end{figure*}
%                                      \         |
%                                        \       |
%                                          \     |
%=======================

%===================================================================
\section{Parameter space and constraints}\label{sec:constr} 
%===================================================================

%------------------------------------------------------------------------------------------
\subsection{Parameter space and DM density}\label{sec:ps}
%------------------------------------------------------------------------------------------

Out of the three free parameters $m_\chi,m_{\tilde t},\lambda_\chi$,
one may be fixed by the condition that the sum of freeze-in and super-WIMP
contributions to the $\chi$ density equals the observed DM abundance.
We choose to fix $\lambda_\chi$ by this condition, and use the
DM mass $m_\chi$ and the mass difference 
\be
\Delta m\equiv m_{\tilde t}-m_\chi
\ee
to describe the remaining two-dimensional parameter space.
Dark matter stability requires $\Delta m>0$. 

In Fig.\,\ref{fig:paramscan1}, we show the resulting coupling $\lambda_\chi$ (green lines)
as function of the DM mass, for fixed $\Delta m=2.5$\,TeV (left) and $\Delta m=100$\,TeV (right).
The freeze-in contribution dominates for $m_\chi\ll m_\chi^{\text{crit}} \simeq 1.6$\, TeV and $40$\,GeV in the two cases, respectively,
corresponding to the ``bulk'' region discussed before. 
The coupling $\lambda_\chi$ required to obtain the measured relic density
increases towards lower DM masses. This can be understood in the following way: as discussed above, for $\Delta m\gg m_\chi$, the freeze-in
yield is approximately independent of $m_\chi$, such that $\Omega h^2 \simeq (\Omega h^2)^{\text{fi}}\propto \lambda_\chi^2m_\chi$.
Requiring $\Omega h^2=0.12$ thus implies $\lambda_\chi\propto \sqrt{m_\chi}$. 

When increasing $m_\chi$ for fixed $\Delta m$, the super-WIMP
contribution becomes larger. Since it is independent of $\lambda_\chi$, its value saturates the constraint $(\Omega h^2)^{\text{sW}}\to 0.12$ for some
finite value of $m_\chi\to m_\chi^{\text{crit}}$. At this point $\lambda_\chi\to 0$, and no solutions providing the measured DM abundance exist for
larger values of $m_\chi$. 
In order to quantify the uncertainty due to the approximate treatment of $2\to 2$ contributions with
threshold enhancement, we show an error band in Fig.\,\ref{fig:paramscan1} around the green line.
The lower boundary corresponds to the result obtained when using a cut parameter $R=1.1$ (see Sec.~\ref{sec:fi}). For the
upper boundary we omit the enhanced $2\to 2$ processes when computing the freeze-in abundance.

\begin{figure*}[t]
\centering
\setlength{\unitlength}{1\textwidth}
\begin{picture}(0.96,0.45)
\put(0.0,-0.012){\includegraphics[trim={80 100 80 80},clip,width=0.96\textwidth]{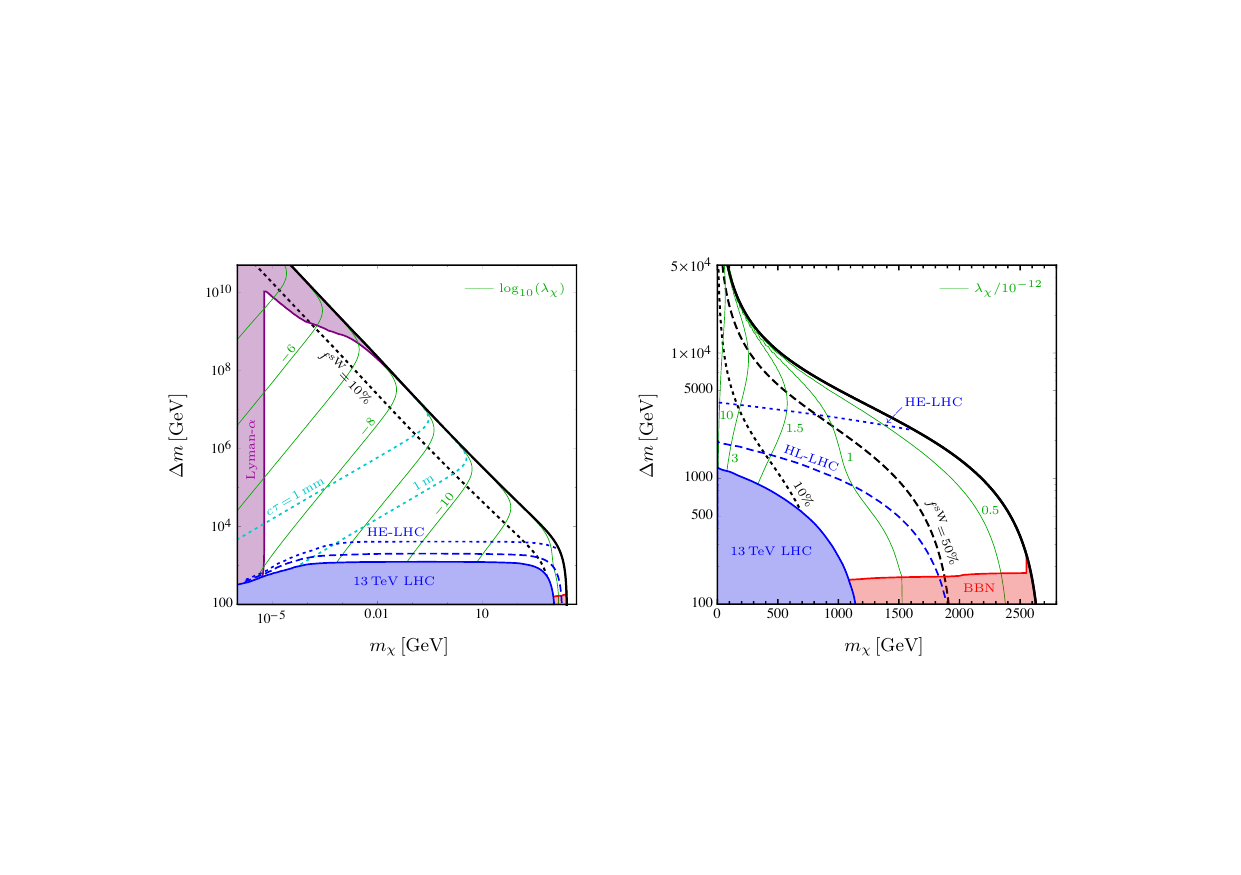}}
\end{picture}
\caption{Cosmologically viable parameter space ($\Omega h^2 \simeq 0.12$) in the $m_\chi$-$\Delta m$
plane displayed logarithmically (left) and linearly (right) in $m_\chi$.
The thin green lines denote contours of constant coupling strength $\lambda_\chi$. 
The black thick line marks the over-closure bound while the black dashed (right panel only) and dotted 
lines denote the contours where $f^\text{sW}=50\%$ and 10\%, respectively. The red and purple
shaded regions are excluded by bounds from BBN and the Lyman-$\alpha$ forest. The blue shaded 
region is excluded by $R$-hadron searches at the 13\,TeV LHC. The dashed and dotted blue curves
denote the respective projection for the 14\,TeV HL-LHC and 27\,TeV HE-LHC, respectively.
}
\label{fig:paramscan2}
\end{figure*}

The full two-dimensional parameter space is shown in Fig.\,\ref{fig:paramscan2}, covering the entire
accessible region of parameter space (left), going up to very large mediator masses, and the
patch for $\Delta m<50$\,TeV, with $m_\chi$ on a linear scale (right), respectively.
The value of $\lambda_\chi$ obtained from imposing the condition $(\Omega h^2)^{\text{fi}}+(\Omega h^2)^{\text{sW}}=0.12$
is indicated by the green contour lines. For the left panel in Fig.\,\ref{fig:paramscan2},
the contours show decades in $\log_{10}\lambda_\chi$, and for the right panel the value of $\lambda_\chi$ normalized to $10^{-12}$. 
In this two-dimensional parameter space, the ``boundary'' corresponds to
the thick black line, for which $(\Omega h^2)^{\text{sW}}\to 0.12$. Values of $(m_\chi,\Delta m)$ above that
line are excluded due to DM overproduction. The region below the black line corresponds to the ``bulk'' as
discussed before. Well inside the ``bulk'', freeze-in production dominates.
The black dashed  (right panel only) and dotted curves show the contour of constant relative super-WIMP contribution $(\Omega h^2)^\text{sW}_\chi/(\Omega h^2)^\text{tot}_\chi=50\%$ and 10\%, respectively.
Below the dotted curve the super-WIMP contribution is subdominant.

In the following we discuss observational signatures that can probe different regions
of the parameter space, including long-lived colored particles at the LHC  ($R$-hadrons)
and during BBN, as well as Lyman-$\alpha$ forest observations.

%=======================
%    \                                           |
%      \                                         |
%        \                                       |
%                                      \         |
%                                        \       |
%                                          \     |
%=======================

%------------------------------------------------------------------------------------------
\subsection{Collider constraints and projections}
%------------------------------------------------------------------------------------------

As discussed in Sec.~\ref{sec:moind} the relic density constraint
implies the existence of long-lived particles in a large
part of the parameter space within the class of models considered here. 
In Figs.~\ref{fig:paramscan1} and~\ref{fig:paramscan2}
we indicate the proper decay length by the cyan dotted contours for $c\tau=1\,$mm
and 1\,m, corresponding to the range in which decays typically take place inside the
detector. Below the latter curve a significant fraction of mediators decay outside the detector. 

The colored mediator $\tilde t$ can be copiously produced at hadron colliders.
For large mediator decay lengths, $c\tau \gtrsim \text{(detector size)}$, 
searches for detector-stable $R$-hadrons provide a promising discovery channel
at the LHC\@. 
Here we constrain the model by current searches at the 13\,TeV LHC
and estimate projections for the HL- and HE-LHC\@. 

Current searches for detector-stable top-squarks with the CMS detector exclude masses up to
$m_{\tilde t}=1250\,$GeV at 95\% CL~\cite{CMS-PAS-EXO-16-036}. This limit is directly applicable to our model
in the region where $c\tau \gg 1\,$m.
For intermediate lifetimes, $c\tau \lesssim 1\,$m, relevant for 
DM masses $m_\chi\lesssim 100\,\text{keV}$, the limit is weakened due to the exponential suppression of the
fraction of decays outside the detector. We use the reinterpretation of the above limit for finite lifetimes
provided in~\cite{Garny:2017rxs} considering the `generic model' for hadronization. The resulting  95\% CL exclusion is
shown in Fig.~\ref{fig:paramscan2} (blue shaded region).
For large DM masses the limit lies entirely in the detector-stable regime and its drop is simply caused by the 
chosen presentation in terms of $m_\chi$ and $\Delta m$. Towards small masses $\Delta m\simeq m_{\tilde t}$ and the drop 
in the limit is due to the exponential suppression of the detector-stable fraction. Still, $R$-hadron searches
constrain the parameter space towards small mediator masses down to the smallest $m_\chi$ consistent 
with Lyman-$\alpha$ bounds, see Sec.~\ref{sec:wdm}.

In order to illustrate the future sensitivity to the model we consider $R$-hadron searches using $3\,\text{ab}^{-1}$ at 14\,TeV (HL-LHC) and $10\,\text{ab}^{-1}$ at 27\,TeV (HE-LHC)\@.
We compute the signal cross sections at the 14 and 27\,TeV with \textsc{NLLFast}~\cite{Beenakker:2010nq} and \textsc{Prospino}~\cite{Beenakker:1997ut}, respectively. 
As the search is based on anomalous ionization loss and time-of-flight the signal efficiencies depend crucially on the
velocity distribution of the produced mediators. To first approximation the velocity distribution stays unchanged 
for constant $m_{\tilde t}/\sqrt{s}$.
We therefore estimate the signal efficiencies by rescaling the ones from~\cite{CMS-PAS-EXO-16-036} (and~\cite{Garny:2017rxs}
for finite lifetimes):
\begin{equation}
({\cal A}\epsilon)_{14\,\text{TeV}}\left(m_{\tilde t}\right) = ({\cal A}\epsilon)_{13\,\text{TeV}}\left(m_{\tilde t}\times {13\,\text{TeV}}/{14\,\text{TeV}}\right)
\end{equation}
and analogous for 27\,TeV.\footnote{Note that a naive use of the recasting of the 8 or 13\,TeV search for heavy stable charged particles
\cite{Khachatryan:2015lla,Heisig:2018kfq} does not resemble this behavior up to $\sqrt{s}=27$\,TeV, but results in significantly smaller signal efficiencies. The reason for this is the decreasing efficiency towards large $p_\text{T}$ for the current CMS detector. We hence implicitly assume an improved performance towards high $p_\text{T}$ for the HE-LHC\@.
}
We estimated the background by rescaling the one reported in~\cite{CMS-PAS-EXO-16-036}
by the cross section ratio $\sigma_{14\,\text{TeV}}/\sigma_{13\,\text{TeV}}$ ($\sigma_{27\,\text{TeV}}/\sigma_{13\,\text{TeV}}$ for 27\,TeV) computed with \textsc{MadGraph5\_aMC@NLO}~\cite{Alwall:2014hca} for the leading background to heavy stable charged particles which is Drell-Yan production of muons. 

In Fig.~\ref{fig:paramscan2} we draw the corresponding projected $95\% \;\text{CL}_\text{s}$-limits for the HL- (blue dashed) and HE-LHC (blue dotted). They reach mediator masses up to 
2000 and 4050\,GeV, respectively (see also the left panel of Fig.~\ref{fig:paramscan2}). The latter can probe the entire DM mass range up to the boundary (where $\Delta m\simeq 1500\,$GeV).

In addition to searches for detector stable objects, in the region $m_\chi\lesssim 100\,\text{keV}$ signatures of
mediators decaying inside the tracker may provide further sensitivity. We expect searches for disappearing 
$R$-hadron tracks and displaced tops to be further promising discovery channels at the HL- and HE-LHC\@.

%------------------------------------------------------------------------------------------
\subsection{BBN bounds}
%------------------------------------------------------------------------------------------

The presence of a metastable colored mediator during the epoch of BBN affects the
predictions for the primordial abundances of light elements through
the energy release from its decay~\cite{Jedamzik:2006xz,Kawasaki:2017bqm} as well as
through bound-state formation with baryonic matter~\cite{Jedamzik:2007qk,Kusakabe:2009jt}.
Due to strong hadro-dissociation processes the former effect is dominant
for a hadronically decaying mediator. We estimate the respective constraints on the parameter space
by applying the results from
\cite{Jedamzik:2006xz} for a hadronic branching ratio of 1 using the mediator freeze-out 
abundance $Y_{\widetilde t}$ and lifetime as computed by \textsc{micrOMEGAS}~5.0.4.
The slight dependence on the mediator mass is approximated by linearly interpolating 
(and extrapolating) the results for 100\,GeV and 1\,TeV in log-log space.
 
The resulting constraints are shown as the red shaded regions  in Figs.~\ref{fig:paramscan1} and~\ref{fig:paramscan2}.
For fixed $m_\chi$ and $\Delta m$, BBN imposes an upper bound on the lifetime, which translates into
a \emph{lower} bound on the coupling $\lambda_\chi$. 
For $m_\chi \ll \Delta m$, both the lifetime and the mediator abundance $Y_{\widetilde t}$ become independent of $m_\chi$, explaining
the almost horizontal exclusion contour in Fig.~\ref{fig:paramscan1}.  
For $\Delta m=2.5\,(100)$\,TeV we find $\lambda_\chi\geq 1\,(6)\times 10^{-14}$ (see Fig.\,\ref{fig:paramscan1}). 
The value of the coupling required for $\Omega h^2=0.12$ is consistent with BBN for most of the parameter space, except for a small strip close to
the ``boundary'', at which $\lambda_\chi\to 0$ (not resolved in Fig.~\ref{fig:paramscan2}), as well as the region $\Delta m\lesssim m_t$, for which the two-body decay $\tilde t\to\chi t$ is
kinematically forbidden, such that the mediator lifetime is increased (see Fig.\,\ref{fig:paramscan2}, right panel).
BBN bounds are stronger than Lyman-$\alpha$ constraints (see below) for $m_\chi\gg$\,keV and $\Delta m\lesssim 10$\,TeV.

%------------------------------------------------------------------------------------------
\subsection{Lyman-$\alpha$ forest bounds}\label{sec:wdm}
%------------------------------------------------------------------------------------------

In this section we consider constraints on the DM and mediator mass from free-streaming
of DM particles, that leads to a suppression of the amplitude of the matter power spectrum
on length scales smaller than the free-streaming scale
\begin{equation}\label{eq:lambdafs}
\lambda_\text{fs}=\int_{0}^{z_\text{prod}} \diff z \frac{v(z)}{H(z)}\,.
\end{equation}
Here $v(z)$ is the typical velocity of DM particles, and $z_\text{prod}$ is the redshift
at which DM is (dominantly) produced.

The power on small scales can be probed by observations of absorption features in the spectra of distant
light sources (quasars) imprinted by intervening clouds of neutral hydrogen, known as Lyman-$\alpha$ forest.
The interpretation of these data depends on various properties of the intergalactic medium (including its redshift-dependent
temperature and adiabatic index) as well as bias parameters that relate the hydrogen distribution to the underlying DM
density field. These astrophysical effects are often described by a number of ``nuisance'' parameters, that need to be varied together
with the cosmological parameters in order to obtain constraints from comparing theoretical predictions based on
hydrodynamical simulations with observations~\cite{Palanque-Delabrouille:2014jca,Palanque-Delabrouille:2015pga}. 
Here we reinterpret such an analysis performed for mixed cold and warm DM~\cite{Baur:2017stq} based on data from BOSS/SDSS~\cite{Palanque-Delabrouille:2013gaa},
XQ-100/VLT \cite{2016A&A...594A..91L}, and MIKE/HIRES~\cite{Viel:2013apy}. 

Since a dedicated analysis is beyond the scope of this work, and in view of astrophysical uncertainties, we estimate Lyman-$\alpha$ constraints on the model considered here by computing the free-streaming length, and comparing to the maximally allowed value taken from~\cite{Baur:2017stq}. More specifically, we translate the $2\sigma$ limits on the
warm DM mass, as function of the warm DM fraction $0\leq f\leq 1$, into an $f$-dependent upper bound $\lambda_\text{fs}^\text{max}(f)$,
and then apply the latter to the model considered here (see below).
For $f=1\,(0.2)$ the analysis of~\cite{Baur:2017stq} yields $m_{\text{WDM}}\geq 4.0\,(1.5)$\,keV, corresponding to 
$\lambda_\text{fs}^\text{max}(1)=0.10$\,Mpc and $\lambda_\text{fs}^\text{max}(0.2)=0.21$\,Mpc (we use 
cosmic parameters from Planck~\cite{Aghanim:2018eyx} for the conversion).

Even though, within the model considered here, DM is composed of a single particle species, the two populations produced via freeze-in and
the super-WIMP mechanism, respectively, feature a different momentum distribution, and therefore different free-steaming lengths, denoted by
$\lambda_\text{fs}^{\text{fi}}$ and $\lambda_\text{fs}^{\text{sW}}$ (see below). We denote the corresponding fractions of the DM density by 
$f^{\text{fi}}=(\Omega_{\chi}h^2)^{\text{fi}}/0.12$ and $f^{\text{sW}}=(\Omega_{\chi}h^2)^{\text{fi}}/0.12$,
such that $f^{\text{fi}}+f^{\text{sW}}=1$ in the cosmologically allowed parameter region.
If for example $\lambda_\text{fs}^{\text{fi}}$ approaches the maximal allowed value, we find that $\lambda_\text{fs}^{\text{sW}}$ is negligibly small
within the majority of the accessible parameter space, such that the fraction of ``warm'' DM is in this case $f^{\text{fi}}$, while the rest
behaves as cold DM on the relevant scales. The same is true vice versa. Therefore, we impose the bound as
\be\label{eq:lyacondition}
  \lambda_\text{fs}^{a} \leq \lambda_\text{fs}^\text{max}(f^a)\,,
\ee
for both $a= \text{fi}, \text{sW}$. This procedure is expected to fail when both production mechanisms produce a comparable free-streaming length, of the order of the
maximally allowed values. In this case the corresponding matter power spectrum can have a more complicated scale-dependence, which requires a dedicated
analysis (see e.g. \cite{Konig:2016dzg,Boulebnane:2017fxw,Garny:2018byk,Murgia:2018now} for related discussions) which is beyond the scope of this work.

Let us now estimate the free-streaming lengths for both production mechanisms.
We assume production is dominated at a redshift interval around some redshift $z_\text{prod}$ 
with a typical momentum $p_\text{prod}$ of the DM particles.
Due to cosmic expansion, the momentum redshifts according to
\begin{equation}
p(z) = p_\text{prod}\frac{1+z}{1+z_\text{prod}}\,,
\end{equation}
and the typical velocity entering \eqref{eq:lambdafs} is given by
\begin{equation}
v(z)=\frac{p(z)}{\sqrt{p(z)^2+m_\chi^2}}\,.
\end{equation}
Together with the standard expression
\begin{equation}
H(z) = H_0 \sqrt{\Omega_\text{m} (1+z)^3 + \Omega_\text{r} (1+z)^4 + \Omega_\Lambda }\,,
\end{equation}
this can be used to compute $\lambda_\text{fs}$ as function of $z_\text{prod}$ and $p_\text{prod}$.
We note that the integral over $z$ is dominated by redshifts $z\gtrsim 10^3$ such that the value
computed with lower integration boundary at $z=0$ and at the redshifts $z\sim 2-5$ relevant 
for Lyman-$\alpha$ observations is practically identical.

For the super-WIMP mechanism, i.e.~late decays $\tilde t\to t\chi$ of the mediator, we assume for the time
of production 
\be
  t_\text{prod}^\text{sW}\simeq 1/\Gamma_\text{decay}\,,
\ee
where $\Gamma_\text{decay}$ is the $\tilde t$ decay rate.
The decay time can be converted to the redshift using
\begin{equation}
t(T)\simeq\frac{0.301}{\sqrt{g_*(T)}}\,\frac{M_\text{Pl}}{ T^2 }\,,
\end{equation}
and then solving $T=(g_{*S}(T)/g_{*S}(T_0))^{1/3}T_0(1+z)$ for the
temperature, where $g_{*(S)}$ are the usual relativistic degrees of freedom.
At the decay time, the kinetic energy of the mediator can be neglected, such that
it decays at rest. The momentum is therefore given by
\begin{equation}
p_\text{prod}^\text{sW}=\frac{\sqrt{m_{\tilde t}^4\!\!\;+m_\chi^4\!\!\;+m_t^4 \!\!\;- 2 (m_{\tilde t}^2m_\chi^2\!\!\; + m_{\tilde t}^2m_t^2\!\!\;+m_\chi^2m_t^2)}}{2 m_{\tilde t}}\,.
\end{equation}
The above equations amount to $v\gamma (z_\text{prod}) \simeq m_{\tilde t}/(2 m_\chi)$ for large $m_{\tilde t}$ where $\gamma=(1-v^2)^{-1/2}$.
Note that for very long mediator lifetimes comparable to the time of recombination, the interaction of the mediator
with baryonic matter can lead to further suppression of the power spectrum~\cite{Sigurdson:2003vy}.
The mediator lifetimes we consider are much smaller than this, even though there may be a tiny region
in parameter space very close to the ``boundary'', for which $\lambda_\chi\to 0$, where this becomes relevant.
We do not consider this possibility further here.

For the freeze-in contribution the dominant part arises before the onset of a (strong) 
Boltzmann suppression of the mediator abundance, that is at $T\sim m_{\tilde t}$. We
approximate the production redshift by
\begin{equation}
z_\text{prod}^\text{fi} \simeq z(T=m_{\tilde t}/x_\text{fi})\,,
\end{equation}
where $x_\text{fi}$ corresponds to the value of $x=m_{\tilde t}/T$ for which the
production rate $\diff Y_\chi/\diff x$ is maximal. For simplicity, we consider production via decay
only for which we find $x_\text{fi}\simeq 2.4$.
Production via scatterings peaks at a value of comparable size, and we therefore expect
the choice adopted here to give a subleading contribution to the error budget.
Note that scatterings are taken into account for the abundance computation.
In order to estimate the typical momentum, we compute the average energy
of DM particles produced
from a thermal distribution of $\tilde t$. As before we focus on production
via decay, yielding $\langle E_\chi \rangle(x_\text{fi})\simeq 0.92 m_{\tilde t}$ for $\Delta m\gg m_t$.

The resulting bounds on the parameter space are shown by the purple-shaded regions in Figs.\,\ref{fig:paramscan1} and  \ref{fig:paramscan2}.
The two distinct exclusion regions in Fig.\,\ref{fig:paramscan1} correspond to the cases 
$\lambda_\text{fs}^{\text{fi}} > \lambda_\text{fs}^\text{max}(f^\text{fi})$
and $\lambda_\text{fs}^{\text{sW}} > \lambda_\text{fs}^\text{max}(f^\text{sW})$, respectively.
In the former case, the free-streaming length becomes large due to the small DM mass, and in the latter case
due to an interplay of the large mass splitting, a long mediator lifetime and a large super-WIMP fraction. 

The resulting exclusion region on $(m_\chi,\Delta m)$ when imposing $\Omega h^2=0.12$
is shown in Fig.\,\ref{fig:paramscan2}.
For small DM masses and $m_{\tilde t}\ll 10^7$\,TeV, the production is dominated by freeze-in (i.e.~$f^{\text{fi}}\simeq 1$).
In this case Lyman-$\alpha$ data impose a lower bound on the DM mass of the order of $m_\chi\gtrsim 6$\,keV (left part of purple-shaded
area in Fig.\,\ref{fig:paramscan2}).
This bound is comparable to warm DM bounds, but slightly stronger, because freeze-in produces a non-thermal spectrum with
slightly higher momentum than in the corresponding thermal case. However, we conservatively attribute an uncertainty of a factor of two
to the precise value, due to the approximate estimate based on the free-streaming length, as well as astrophysical uncertainties (see above).

For the region in parameter space for which the super-WIMP mechanism gives a sizeable contribution, Lyman-$\alpha$ data exclude
large mediator masses, and require e.g. $m_{\tilde t}\lesssim 10^6$\,TeV for $m_\chi\simeq$\,MeV (right part of purple-shaded
area in Fig.\,\ref{fig:paramscan2}).
The decay of mediators above this bound would lead to a too large free-streaming length, because
the heavy mediator converts its rest mass into kinetic energy of the DM particles. 
For $m_{\tilde t}\gtrsim 10^7$\,TeV this excludes points in parameter space with a
fraction of DM produced via the super-WIMP mechanism down to $f^{\text{sW}}\gtrsim 10\%$. 

The exclusion contour in Fig.\,\ref{fig:paramscan2} close to the crossing point of the two regions discussed above should
be regarded as conservative, because both populations have sizeable free-streaming length in that part of parameter space.
The approach outlined above tends to underestimate the suppression of the power spectrum in this case.
As mentioned above, a proper treatment of this region goes beyond the scope of this work.

%===================================================================
\section{Conclusion}\label{sec:summary} 
%===================================================================

In this work we studied a class of DM models comprising a $Z_2$-odd dark sector
that contains a feebly interacting DM particle along with a mediator that transforms
non-trivially under the SM gauge group. The DM particle never thermalizes and
is produced via a combination of freeze-in on the one hand, and late decays of frozen-out
mediator particles, known as super-WIMP mechanism,  on the other hand.
Despite the fact that DM interactions with the SM are tiny -- well in agreement with
null-searches in direct and indirect detection experiments -- the presence of the mediator
leads to characteristic signatures that can be probed by searches for long-lived particles
at colliders, via Lyman-$\alpha$ forest observations, as well as the determination of
primordial element abundances.

The interplay of freeze-in and super-WIMP production leads to a finite region
in parameter space for which the observed DM abundance can be explained. Taking
a Majorana DM particle $\chi$ and a colored top-philic scalar mediator ${\tilde t}$ as an example,
the DM mass is bounded by $m_\chi\leq m_\chi^\text{max}\simeq2700$\,GeV. In addition, there is a maximal
possible mediator mass, depending on the value of $m_\chi$. For example, 
$m_{\tilde t}\leq m_{\text{med}}^{\text{max}}\simeq (6,\,3\times10^3,\,2\times10^6)$\,TeV for $m_\chi=(10^3,\,1,\,10^{-3})$\,GeV.
In addition, we highlight a simple parametric relation between the mediator lifetime and the masses.
For DM mass in the MeV range and mediator decay length on detector scales (mm--m), the mediator mass
is in the (multi-)TeV region.

We explore experimental probes within the entire accessible parameter space of DM and mediator masses,
and provide up-to-date exclusion limits. 
We find that the parameter space is constrained from all sides.
Mediator masses around the TeV-scale and below are excluded from $R$-hadron searches at the $13$\,TeV LHC for $m_\chi\lesssim 1$\,TeV,
and by BBN for $m_\chi\gtrsim 1$\,TeV.
On the other hand, very heavy mediators with mass above $10^6$--$10^7$\,TeV are in conflict with recent
Lyman-$\alpha$ forest observations, as are DM masses below about $6\,$keV. In addition, a tiny region
close to the boundary of the accessible parameter space, with $m_{\tilde t}\lesssim m_{\tilde t}^{\text{max}}$, 
is excluded by BBN and Lyman-$\alpha$ observations for $m_{\tilde t}$ below and above $10$\,TeV, respectively
 (not fully resolved in Fig.\,\ref{fig:paramscan2}). In this region a large fraction of the DM abundance is
produced via the super-WIMP mechanism, and the mediator lifetime becomes particularly large.
We also provide projections for high luminosity and $27$\,TeV high energy upgrades of the LHC, that are sensitive to mediator
masses up to around $2$ and $4$\,TeV, respectively.

While we focus our phenomenological analysis in this work on a specific simplified model, we emphasize that the
qualitative features remain the same in general. For example, for a lepto-philic mediator $m_\chi^\text{max}$
and $m_{\text{med}}^{\text{max}}$ would be smaller due to its larger freeze-out abundance, while the collider bound would
be weaker due to the reduced production cross section. In the future, it would be interesting to perform a dedicated 
analysis of Lyman-$\alpha$ forest bounds for combined super-WIMP and freeze-in production.

%===================================================================
\section*{Acknowledgements}
%===================================================================

We thank Andreas Goudelis and Alexander Pukhov
for very helpful discussions. 
We acknowledge support by the German Research Foundation (DFG) through the 
research unit ``New physics at the LHC''.

%%%  Bibliography 
\bibliography{bibliography}{}

\begin{thebibliography}{50}
\expandafter\ifx\csname natexlab\endcsname\relax\def\natexlab#1{#1}\fi
\expandafter\ifx\csname bibnamefont\endcsname\relax
  \def\bibnamefont#1{#1}\fi
\expandafter\ifx\csname bibfnamefont\endcsname\relax
  \def\bibfnamefont#1{#1}\fi
\expandafter\ifx\csname citenamefont\endcsname\relax
  \def\citenamefont#1{#1}\fi
\expandafter\ifx\csname url\endcsname\relax
  \def\url#1{\texttt{#1}}\fi
\expandafter\ifx\csname urlprefix\endcsname\relax\def\urlprefix{URL }\fi
\providecommand{\bibinfo}[2]{#2}
\providecommand{\eprint}[2][]{\url{#2}}

\bibitem[{\citenamefont{Bolz et~al.}(2001)\citenamefont{Bolz, Brandenburg, and
  Buchm{\"u}ller}}]{Bolz:2000fu}
\bibinfo{author}{\bibfnamefont{M.}~\bibnamefont{Bolz}},
  \bibinfo{author}{\bibfnamefont{A.}~\bibnamefont{Brandenburg}},
  \bibnamefont{and}
  \bibinfo{author}{\bibfnamefont{W.}~\bibnamefont{Buchm{\"u}ller}},
  \bibinfo{journal}{Nucl. Phys.} \textbf{\bibinfo{volume}{B606}},
  \bibinfo{pages}{518} (\bibinfo{year}{2001}), \bibinfo{note}{[Erratum: Nucl.
  Phys. B790, 336 (2008)]}, \eprint{hep-ph/0012052}.

\bibitem[{\citenamefont{Pradler and Steffen}(2007)}]{Pradler:2006hh}
\bibinfo{author}{\bibfnamefont{J.}~\bibnamefont{Pradler}} \bibnamefont{and}
  \bibinfo{author}{\bibfnamefont{F.~D.} \bibnamefont{Steffen}},
  \bibinfo{journal}{Phys. Lett.} \textbf{\bibinfo{volume}{B648}},
  \bibinfo{pages}{224} (\bibinfo{year}{2007}), \eprint{hep-ph/0612291}.

\bibitem[{\citenamefont{McDonald}(2002)}]{McDonald:2001vt}
\bibinfo{author}{\bibfnamefont{J.}~\bibnamefont{McDonald}},
  \bibinfo{journal}{Phys. Rev. Lett.} \textbf{\bibinfo{volume}{88}},
  \bibinfo{pages}{091304} (\bibinfo{year}{2002}), \eprint{hep-ph/0106249}.

\bibitem[{\citenamefont{Covi et~al.}(2002)\citenamefont{Covi, Roszkowski, and
  Small}}]{Covi:2002vw}
\bibinfo{author}{\bibfnamefont{L.}~\bibnamefont{Covi}},
  \bibinfo{author}{\bibfnamefont{L.}~\bibnamefont{Roszkowski}},
  \bibnamefont{and} \bibinfo{author}{\bibfnamefont{M.}~\bibnamefont{Small}},
  \bibinfo{journal}{JHEP} \textbf{\bibinfo{volume}{07}}, \bibinfo{pages}{023}
  (\bibinfo{year}{2002}), \eprint{hep-ph/0206119}.

\bibitem[{\citenamefont{Asaka et~al.}(2006)\citenamefont{Asaka, Ishiwata, and
  Moroi}}]{Asaka:2005cn}
\bibinfo{author}{\bibfnamefont{T.}~\bibnamefont{Asaka}},
  \bibinfo{author}{\bibfnamefont{K.}~\bibnamefont{Ishiwata}}, \bibnamefont{and}
  \bibinfo{author}{\bibfnamefont{T.}~\bibnamefont{Moroi}},
  \bibinfo{journal}{Phys. Rev.} \textbf{\bibinfo{volume}{D73}},
  \bibinfo{pages}{051301} (\bibinfo{year}{2006}), \eprint{hep-ph/0512118}.

\bibitem[{\citenamefont{Hall et~al.}(2010)\citenamefont{Hall, Jedamzik,
  March-Russell, and West}}]{Hall:2009bx}
\bibinfo{author}{\bibfnamefont{L.~J.} \bibnamefont{Hall}},
  \bibinfo{author}{\bibfnamefont{K.}~\bibnamefont{Jedamzik}},
  \bibinfo{author}{\bibfnamefont{J.}~\bibnamefont{March-Russell}},
  \bibnamefont{and} \bibinfo{author}{\bibfnamefont{S.~M.} \bibnamefont{West}},
  \bibinfo{journal}{JHEP} \textbf{\bibinfo{volume}{03}}, \bibinfo{pages}{080}
  (\bibinfo{year}{2010}), \eprint{0911.1120}.

\bibitem[{\citenamefont{Covi et~al.}(1999)\citenamefont{Covi, Kim, and
  Roszkowski}}]{Covi:1999ty}
\bibinfo{author}{\bibfnamefont{L.}~\bibnamefont{Covi}},
  \bibinfo{author}{\bibfnamefont{J.~E.} \bibnamefont{Kim}}, \bibnamefont{and}
  \bibinfo{author}{\bibfnamefont{L.}~\bibnamefont{Roszkowski}},
  \bibinfo{journal}{Phys. Rev. Lett.} \textbf{\bibinfo{volume}{82}},
  \bibinfo{pages}{4180} (\bibinfo{year}{1999}), \eprint{hep-ph/9905212}.

\bibitem[{\citenamefont{Feng et~al.}(2003)\citenamefont{Feng, Rajaraman, and
  Takayama}}]{Feng:2003uy}
\bibinfo{author}{\bibfnamefont{J.~L.} \bibnamefont{Feng}},
  \bibinfo{author}{\bibfnamefont{A.}~\bibnamefont{Rajaraman}},
  \bibnamefont{and} \bibinfo{author}{\bibfnamefont{F.}~\bibnamefont{Takayama}},
  \bibinfo{journal}{Phys. Rev.} \textbf{\bibinfo{volume}{D68}},
  \bibinfo{pages}{063504} (\bibinfo{year}{2003}), \eprint{hep-ph/0306024}.

\bibitem[{\citenamefont{Palanque-Delabrouille
  et~al.}(2013)}]{Palanque-Delabrouille:2013gaa}
\bibinfo{author}{\bibfnamefont{N.}~\bibnamefont{Palanque-Delabrouille}}
  \bibnamefont{et~al.}, \bibinfo{journal}{Astron. Astrophys.}
  \textbf{\bibinfo{volume}{559}}, \bibinfo{pages}{A85} (\bibinfo{year}{2013}),
  \eprint{1306.5896}.

\bibitem[{\citenamefont{{L{\'o}pez} et~al.}(2016)\citenamefont{{L{\'o}pez},
  {D'Odorico}, {Ellison}, {Becker}, {Christensen}, {Cupani}, {Denney},
  {P{\^a}ris}, {Worseck}, {Berg} et~al.}}]{2016A&A...594A..91L}
\bibinfo{author}{\bibfnamefont{S.}~\bibnamefont{{L{\'o}pez}}},
  \bibinfo{author}{\bibfnamefont{V.}~\bibnamefont{{D'Odorico}}},
  \bibinfo{author}{\bibfnamefont{S.~L.} \bibnamefont{{Ellison}}},
  \bibinfo{author}{\bibfnamefont{G.~D.} \bibnamefont{{Becker}}},
  \bibinfo{author}{\bibfnamefont{L.}~\bibnamefont{{Christensen}}},
  \bibinfo{author}{\bibfnamefont{G.}~\bibnamefont{{Cupani}}},
  \bibinfo{author}{\bibfnamefont{K.~D.} \bibnamefont{{Denney}}},
  \bibinfo{author}{\bibfnamefont{I.}~\bibnamefont{{P{\^a}ris}}},
  \bibinfo{author}{\bibfnamefont{G.}~\bibnamefont{{Worseck}}},
  \bibinfo{author}{\bibfnamefont{T.~A.~M.} \bibnamefont{{Berg}}},
  \bibnamefont{et~al.}, \bibinfo{journal}{\aap} \textbf{\bibinfo{volume}{594}},
  \bibinfo{eid}{A91} (\bibinfo{year}{2016}), \eprint{1607.08776}.

\bibitem[{\citenamefont{Viel et~al.}(2013)\citenamefont{Viel, Becker, Bolton,
  and Haehnelt}}]{Viel:2013apy}
\bibinfo{author}{\bibfnamefont{M.}~\bibnamefont{Viel}},
  \bibinfo{author}{\bibfnamefont{G.~D.} \bibnamefont{Becker}},
  \bibinfo{author}{\bibfnamefont{J.~S.} \bibnamefont{Bolton}},
  \bibnamefont{and} \bibinfo{author}{\bibfnamefont{M.~G.}
  \bibnamefont{Haehnelt}}, \bibinfo{journal}{Phys. Rev.}
  \textbf{\bibinfo{volume}{D88}}, \bibinfo{pages}{043502}
  (\bibinfo{year}{2013}), \eprint{1306.2314}.

\bibitem[{\citenamefont{Ibarra et~al.}(2015)\citenamefont{Ibarra, Pierce, Shah,
  and Vogl}}]{Ibarra:2015nca}
\bibinfo{author}{\bibfnamefont{A.}~\bibnamefont{Ibarra}},
  \bibinfo{author}{\bibfnamefont{A.}~\bibnamefont{Pierce}},
  \bibinfo{author}{\bibfnamefont{N.~R.} \bibnamefont{Shah}}, \bibnamefont{and}
  \bibinfo{author}{\bibfnamefont{S.}~\bibnamefont{Vogl}},
  \bibinfo{journal}{Phys. Rev.} \textbf{\bibinfo{volume}{D91}},
  \bibinfo{pages}{095018} (\bibinfo{year}{2015}), \eprint{1501.03164}.

\bibitem[{\citenamefont{Garny et~al.}(2018{\natexlab{a}})\citenamefont{Garny,
  Heisig, Hufnagel, and L{\"u}lf}}]{Garny:2018icg}
\bibinfo{author}{\bibfnamefont{M.}~\bibnamefont{Garny}},
  \bibinfo{author}{\bibfnamefont{J.}~\bibnamefont{Heisig}},
  \bibinfo{author}{\bibfnamefont{M.}~\bibnamefont{Hufnagel}}, \bibnamefont{and}
  \bibinfo{author}{\bibfnamefont{B.}~\bibnamefont{L{\"u}lf}},
  \bibinfo{journal}{Phys. Rev.} \textbf{\bibinfo{volume}{D97}},
  \bibinfo{pages}{075002} (\bibinfo{year}{2018}{\natexlab{a}}),
  \eprint{1802.00814}.

\bibitem[{\citenamefont{Ibarra et~al.}(2009)\citenamefont{Ibarra, Ringwald, and
  Weniger}}]{Ibarra:2008kn}
\bibinfo{author}{\bibfnamefont{A.}~\bibnamefont{Ibarra}},
  \bibinfo{author}{\bibfnamefont{A.}~\bibnamefont{Ringwald}}, \bibnamefont{and}
  \bibinfo{author}{\bibfnamefont{C.}~\bibnamefont{Weniger}},
  \bibinfo{journal}{JCAP} \textbf{\bibinfo{volume}{0901}}, \bibinfo{pages}{003}
  (\bibinfo{year}{2009}), \eprint{0809.3196}.

\bibitem[{\citenamefont{De~Simone et~al.}(2010)\citenamefont{De~Simone, Garny,
  Ibarra, and Weniger}}]{DeSimone:2010tr}
\bibinfo{author}{\bibfnamefont{A.}~\bibnamefont{De~Simone}},
  \bibinfo{author}{\bibfnamefont{M.}~\bibnamefont{Garny}},
  \bibinfo{author}{\bibfnamefont{A.}~\bibnamefont{Ibarra}}, \bibnamefont{and}
  \bibinfo{author}{\bibfnamefont{C.}~\bibnamefont{Weniger}},
  \bibinfo{journal}{JCAP} \textbf{\bibinfo{volume}{1007}}, \bibinfo{pages}{017}
  (\bibinfo{year}{2010}), \eprint{1004.4890}.

\bibitem[{\citenamefont{Covi and Dradi}(2014)}]{Covi:2014fba}
\bibinfo{author}{\bibfnamefont{L.}~\bibnamefont{Covi}} \bibnamefont{and}
  \bibinfo{author}{\bibfnamefont{F.}~\bibnamefont{Dradi}},
  \bibinfo{journal}{JCAP} \textbf{\bibinfo{volume}{1410}}, \bibinfo{pages}{039}
  (\bibinfo{year}{2014}), \eprint{1403.4923}.

\bibitem[{\citenamefont{Arcadi and Covi}(2013)}]{Arcadi:2013aba}
\bibinfo{author}{\bibfnamefont{G.}~\bibnamefont{Arcadi}} \bibnamefont{and}
  \bibinfo{author}{\bibfnamefont{L.}~\bibnamefont{Covi}},
  \bibinfo{journal}{JCAP} \textbf{\bibinfo{volume}{1308}}, \bibinfo{pages}{005}
  (\bibinfo{year}{2013}), \eprint{1305.6587}.

\bibitem[{\citenamefont{Arcadi et~al.}(2014)\citenamefont{Arcadi, Covi, and
  Dradi}}]{Arcadi:2014tsa}
\bibinfo{author}{\bibfnamefont{G.}~\bibnamefont{Arcadi}},
  \bibinfo{author}{\bibfnamefont{L.}~\bibnamefont{Covi}}, \bibnamefont{and}
  \bibinfo{author}{\bibfnamefont{F.}~\bibnamefont{Dradi}},
  \bibinfo{journal}{JCAP} \textbf{\bibinfo{volume}{1410}}, \bibinfo{pages}{063}
  (\bibinfo{year}{2014}), \eprint{1408.1005}.

\bibitem[{\citenamefont{B{\'e}langer et~al.}(2018)\citenamefont{B{\'e}langer,
  Boudjema, Goudelis, Pukhov, and Zaldivar}}]{Belanger:2018ccd}
\bibinfo{author}{\bibfnamefont{G.}~\bibnamefont{B{\'e}langer}},
  \bibinfo{author}{\bibfnamefont{F.}~\bibnamefont{Boudjema}},
  \bibinfo{author}{\bibfnamefont{A.}~\bibnamefont{Goudelis}},
  \bibinfo{author}{\bibfnamefont{A.}~\bibnamefont{Pukhov}}, \bibnamefont{and}
  \bibinfo{author}{\bibfnamefont{B.}~\bibnamefont{Zaldivar}},
  \bibinfo{journal}{Comput. Phys. Commun.} \textbf{\bibinfo{volume}{231}},
  \bibinfo{pages}{173} (\bibinfo{year}{2018}), \eprint{1801.03509}.

\bibitem[{\citenamefont{Garny et~al.}(2017)\citenamefont{Garny, Heisig,
  L{\"u}lf, and Vogl}}]{Garny:2017rxs}
\bibinfo{author}{\bibfnamefont{M.}~\bibnamefont{Garny}},
  \bibinfo{author}{\bibfnamefont{J.}~\bibnamefont{Heisig}},
  \bibinfo{author}{\bibfnamefont{B.}~\bibnamefont{L{\"u}lf}}, \bibnamefont{and}
  \bibinfo{author}{\bibfnamefont{S.}~\bibnamefont{Vogl}},
  \bibinfo{journal}{Phys. Rev.} \textbf{\bibinfo{volume}{D96}},
  \bibinfo{pages}{103521} (\bibinfo{year}{2017}), \eprint{1705.09292}.

\bibitem[{\citenamefont{Liew and Luo}(2017)}]{Liew:2016hqo}
\bibinfo{author}{\bibfnamefont{S.~P.} \bibnamefont{Liew}} \bibnamefont{and}
  \bibinfo{author}{\bibfnamefont{F.}~\bibnamefont{Luo}},
  \bibinfo{journal}{JHEP} \textbf{\bibinfo{volume}{02}}, \bibinfo{pages}{091}
  (\bibinfo{year}{2017}), \eprint{1611.08133}.

\bibitem[{\citenamefont{Mitridate et~al.}(2017)\citenamefont{Mitridate, Redi,
  Smirnov, and Strumia}}]{Mitridate:2017izz}
\bibinfo{author}{\bibfnamefont{A.}~\bibnamefont{Mitridate}},
  \bibinfo{author}{\bibfnamefont{M.}~\bibnamefont{Redi}},
  \bibinfo{author}{\bibfnamefont{J.}~\bibnamefont{Smirnov}}, \bibnamefont{and}
  \bibinfo{author}{\bibfnamefont{A.}~\bibnamefont{Strumia}},
  \bibinfo{journal}{JCAP} \textbf{\bibinfo{volume}{1705}}, \bibinfo{pages}{006}
  (\bibinfo{year}{2017}), \eprint{1702.01141}.

\bibitem[{\citenamefont{Biondini and Laine}(2018)}]{Biondini:2018pwp}
\bibinfo{author}{\bibfnamefont{S.}~\bibnamefont{Biondini}} \bibnamefont{and}
  \bibinfo{author}{\bibfnamefont{M.}~\bibnamefont{Laine}},
  \bibinfo{journal}{JHEP} \textbf{\bibinfo{volume}{04}}, \bibinfo{pages}{072}
  (\bibinfo{year}{2018}), \eprint{1801.05821}.

\bibitem[{\citenamefont{Heisig}(2018)}]{Heisig:2018teh}
\bibinfo{author}{\bibfnamefont{J.}~\bibnamefont{Heisig}}, in
  \emph{\bibinfo{booktitle}{{53rd Rencontres de Moriond on Electroweak
  Interactions and Unified Theories (Moriond EW 2018) La Thuile, Italy, March
  10-17, 2018}}} (\bibinfo{year}{2018}), \eprint{1805.07361}.

\bibitem[{\citenamefont{Bernal et~al.}(2017)\citenamefont{Bernal, Heikinheimo,
  Tenkanen, Tuominen, and Vaskonen}}]{Bernal:2017kxu}
\bibinfo{author}{\bibfnamefont{N.}~\bibnamefont{Bernal}},
  \bibinfo{author}{\bibfnamefont{M.}~\bibnamefont{Heikinheimo}},
  \bibinfo{author}{\bibfnamefont{T.}~\bibnamefont{Tenkanen}},
  \bibinfo{author}{\bibfnamefont{K.}~\bibnamefont{Tuominen}}, \bibnamefont{and}
  \bibinfo{author}{\bibfnamefont{V.}~\bibnamefont{Vaskonen}},
  \bibinfo{journal}{Int. J. Mod. Phys.} \textbf{\bibinfo{volume}{A32}},
  \bibinfo{pages}{1730023} (\bibinfo{year}{2017}), \eprint{1706.07442}.

\bibitem[{\citenamefont{Co et~al.}(2015)\citenamefont{Co, D'Eramo, Hall, and
  Pappadopulo}}]{Co:2015pka}
\bibinfo{author}{\bibfnamefont{R.~T.} \bibnamefont{Co}},
  \bibinfo{author}{\bibfnamefont{F.}~\bibnamefont{D'Eramo}},
  \bibinfo{author}{\bibfnamefont{L.~J.} \bibnamefont{Hall}}, \bibnamefont{and}
  \bibinfo{author}{\bibfnamefont{D.}~\bibnamefont{Pappadopulo}},
  \bibinfo{journal}{JCAP} \textbf{\bibinfo{volume}{1512}}, \bibinfo{pages}{024}
  (\bibinfo{year}{2015}), \eprint{1506.07532}.

\bibitem[{\citenamefont{Hessler et~al.}(2017)\citenamefont{Hessler, Ibarra,
  Molinaro, and Vogl}}]{Hessler:2016kwm}
\bibinfo{author}{\bibfnamefont{A.~G.} \bibnamefont{Hessler}},
  \bibinfo{author}{\bibfnamefont{A.}~\bibnamefont{Ibarra}},
  \bibinfo{author}{\bibfnamefont{E.}~\bibnamefont{Molinaro}}, \bibnamefont{and}
  \bibinfo{author}{\bibfnamefont{S.}~\bibnamefont{Vogl}},
  \bibinfo{journal}{JHEP} \textbf{\bibinfo{volume}{01}}, \bibinfo{pages}{100}
  (\bibinfo{year}{2017}), \eprint{1611.09540}.

\bibitem[{\citenamefont{Ghosh et~al.}(2017)\citenamefont{Ghosh, Mondal, and
  Mukhopadhyaya}}]{Ghosh:2017vhe}
\bibinfo{author}{\bibfnamefont{A.}~\bibnamefont{Ghosh}},
  \bibinfo{author}{\bibfnamefont{T.}~\bibnamefont{Mondal}}, \bibnamefont{and}
  \bibinfo{author}{\bibfnamefont{B.}~\bibnamefont{Mukhopadhyaya}},
  \bibinfo{journal}{JHEP} \textbf{\bibinfo{volume}{12}}, \bibinfo{pages}{136}
  (\bibinfo{year}{2017}), \eprint{1706.06815}.

\bibitem[{\citenamefont{D'Eramo et~al.}(2018)\citenamefont{D'Eramo, Fernandez,
  and Profumo}}]{DEramo:2017ecx}
\bibinfo{author}{\bibfnamefont{F.}~\bibnamefont{D'Eramo}},
  \bibinfo{author}{\bibfnamefont{N.}~\bibnamefont{Fernandez}},
  \bibnamefont{and} \bibinfo{author}{\bibfnamefont{S.}~\bibnamefont{Profumo}},
  \bibinfo{journal}{JCAP} \textbf{\bibinfo{volume}{1802}}, \bibinfo{pages}{046}
  (\bibinfo{year}{2018}), \eprint{1712.07453}.

\bibitem[{\citenamefont{Brooijmans et~al.}(2018)}]{Brooijmans:2018xbu}
\bibinfo{author}{\bibfnamefont{G.}~\bibnamefont{Brooijmans}}
  \bibnamefont{et~al.}, in \emph{\bibinfo{booktitle}{{10th Les Houches Workshop
  on Physics at TeV Colliders (PhysTeV 2017) Les Houches, France, June 5-23,
  2017}}} (\bibinfo{year}{2018}), \eprint{1803.10379}.

\bibitem[{\citenamefont{Calibbi et~al.}(2018)\citenamefont{Calibbi,
  Lopez-Honorez, Lowette, and Mariotti}}]{Calibbi:2018fqf}
\bibinfo{author}{\bibfnamefont{L.}~\bibnamefont{Calibbi}},
  \bibinfo{author}{\bibfnamefont{L.}~\bibnamefont{Lopez-Honorez}},
  \bibinfo{author}{\bibfnamefont{S.}~\bibnamefont{Lowette}}, \bibnamefont{and}
  \bibinfo{author}{\bibfnamefont{A.}~\bibnamefont{Mariotti}},
  \bibinfo{journal}{JHEP} \textbf{\bibinfo{volume}{09}}, \bibinfo{pages}{037}
  (\bibinfo{year}{2018}), \eprint{1805.04423}.

\bibitem[{\citenamefont{Khachatryan et~al.}(2016)}]{CMS-PAS-EXO-16-036}
\bibinfo{author}{\bibfnamefont{V.}~\bibnamefont{Khachatryan}}
  \bibnamefont{et~al.} (\bibinfo{collaboration}{CMS}), \bibinfo{type}{Tech.
  Rep.} \bibinfo{number}{CMS-PAS-EXO-16-036} (\bibinfo{year}{2016}),
  \urlprefix\url{http://cds.cern.ch/record/2205281}.

\bibitem[{\citenamefont{Beenakker et~al.}(2010)\citenamefont{Beenakker,
  Brensing, Kr{\"a}mer, Kulesza, Laenen, and Niessen}}]{Beenakker:2010nq}
\bibinfo{author}{\bibfnamefont{W.}~\bibnamefont{Beenakker}},
  \bibinfo{author}{\bibfnamefont{S.}~\bibnamefont{Brensing}},
  \bibinfo{author}{\bibfnamefont{M.}~\bibnamefont{Kr{\"a}mer}},
  \bibinfo{author}{\bibfnamefont{A.}~\bibnamefont{Kulesza}},
  \bibinfo{author}{\bibfnamefont{E.}~\bibnamefont{Laenen}}, \bibnamefont{and}
  \bibinfo{author}{\bibfnamefont{I.}~\bibnamefont{Niessen}},
  \bibinfo{journal}{JHEP} \textbf{\bibinfo{volume}{08}}, \bibinfo{pages}{098}
  (\bibinfo{year}{2010}), \eprint{1006.4771}.

\bibitem[{\citenamefont{Beenakker et~al.}(1998)\citenamefont{Beenakker,
  Kr{\"a}mer, Plehn, Spira, and Zerwas}}]{Beenakker:1997ut}
\bibinfo{author}{\bibfnamefont{W.}~\bibnamefont{Beenakker}},
  \bibinfo{author}{\bibfnamefont{M.}~\bibnamefont{Kr{\"a}mer}},
  \bibinfo{author}{\bibfnamefont{T.}~\bibnamefont{Plehn}},
  \bibinfo{author}{\bibfnamefont{M.}~\bibnamefont{Spira}}, \bibnamefont{and}
  \bibinfo{author}{\bibfnamefont{P.~M.} \bibnamefont{Zerwas}},
  \bibinfo{journal}{Nucl. Phys.} \textbf{\bibinfo{volume}{B515}},
  \bibinfo{pages}{3} (\bibinfo{year}{1998}), \eprint{hep-ph/9710451}.

\bibitem[{\citenamefont{Khachatryan et~al.}(2015)}]{Khachatryan:2015lla}
\bibinfo{author}{\bibfnamefont{V.}~\bibnamefont{Khachatryan}}
  \bibnamefont{et~al.} (\bibinfo{collaboration}{CMS}), \bibinfo{journal}{Eur.
  Phys. J.} \textbf{\bibinfo{volume}{C75}}, \bibinfo{pages}{325}
  (\bibinfo{year}{2015}), \eprint{1502.02522}.

\bibitem[{\citenamefont{Heisig et~al.}(2019)\citenamefont{Heisig, Kraml, and
  Lessa}}]{Heisig:2018kfq}
\bibinfo{author}{\bibfnamefont{J.}~\bibnamefont{Heisig}},
  \bibinfo{author}{\bibfnamefont{S.}~\bibnamefont{Kraml}}, \bibnamefont{and}
  \bibinfo{author}{\bibfnamefont{A.}~\bibnamefont{Lessa}},
  \bibinfo{journal}{Phys. Lett.} \textbf{\bibinfo{volume}{B788}},
  \bibinfo{pages}{87} (\bibinfo{year}{2019}), \eprint{1808.05229}.

\bibitem[{\citenamefont{Alwall et~al.}(2014)\citenamefont{Alwall, Frederix,
  Frixione, Hirschi, Maltoni, Mattelaer, Shao, Stelzer, Torrielli, and
  Zaro}}]{Alwall:2014hca}
\bibinfo{author}{\bibfnamefont{J.}~\bibnamefont{Alwall}},
  \bibinfo{author}{\bibfnamefont{R.}~\bibnamefont{Frederix}},
  \bibinfo{author}{\bibfnamefont{S.}~\bibnamefont{Frixione}},
  \bibinfo{author}{\bibfnamefont{V.}~\bibnamefont{Hirschi}},
  \bibinfo{author}{\bibfnamefont{F.}~\bibnamefont{Maltoni}},
  \bibinfo{author}{\bibfnamefont{O.}~\bibnamefont{Mattelaer}},
  \bibinfo{author}{\bibfnamefont{H.~S.} \bibnamefont{Shao}},
  \bibinfo{author}{\bibfnamefont{T.}~\bibnamefont{Stelzer}},
  \bibinfo{author}{\bibfnamefont{P.}~\bibnamefont{Torrielli}},
  \bibnamefont{and} \bibinfo{author}{\bibfnamefont{M.}~\bibnamefont{Zaro}},
  \bibinfo{journal}{JHEP} \textbf{\bibinfo{volume}{07}}, \bibinfo{pages}{079}
  (\bibinfo{year}{2014}), \eprint{1405.0301}.

\bibitem[{\citenamefont{Jedamzik}(2006)}]{Jedamzik:2006xz}
\bibinfo{author}{\bibfnamefont{K.}~\bibnamefont{Jedamzik}},
  \bibinfo{journal}{Phys. Rev.} \textbf{\bibinfo{volume}{D74}},
  \bibinfo{pages}{103509} (\bibinfo{year}{2006}), \eprint{hep-ph/0604251}.

\bibitem[{\citenamefont{Kawasaki et~al.}(2018)\citenamefont{Kawasaki, Kohri,
  Moroi, and Takaesu}}]{Kawasaki:2017bqm}
\bibinfo{author}{\bibfnamefont{M.}~\bibnamefont{Kawasaki}},
  \bibinfo{author}{\bibfnamefont{K.}~\bibnamefont{Kohri}},
  \bibinfo{author}{\bibfnamefont{T.}~\bibnamefont{Moroi}}, \bibnamefont{and}
  \bibinfo{author}{\bibfnamefont{Y.}~\bibnamefont{Takaesu}},
  \bibinfo{journal}{Phys. Rev.} \textbf{\bibinfo{volume}{D97}},
  \bibinfo{pages}{023502} (\bibinfo{year}{2018}), \eprint{1709.01211}.

\bibitem[{\citenamefont{Jedamzik}(2008)}]{Jedamzik:2007qk}
\bibinfo{author}{\bibfnamefont{K.}~\bibnamefont{Jedamzik}},
  \bibinfo{journal}{JCAP} \textbf{\bibinfo{volume}{0803}}, \bibinfo{pages}{008}
  (\bibinfo{year}{2008}), \eprint{0710.5153}.

\bibitem[{\citenamefont{Kusakabe et~al.}(2009)\citenamefont{Kusakabe, Kajino,
  Yoshida, and Mathews}}]{Kusakabe:2009jt}
\bibinfo{author}{\bibfnamefont{M.}~\bibnamefont{Kusakabe}},
  \bibinfo{author}{\bibfnamefont{T.}~\bibnamefont{Kajino}},
  \bibinfo{author}{\bibfnamefont{T.}~\bibnamefont{Yoshida}}, \bibnamefont{and}
  \bibinfo{author}{\bibfnamefont{G.~J.} \bibnamefont{Mathews}},
  \bibinfo{journal}{Phys. Rev.} \textbf{\bibinfo{volume}{D80}},
  \bibinfo{pages}{103501} (\bibinfo{year}{2009}), \eprint{0906.3516}.

\bibitem[{\citenamefont{Palanque-Delabrouille
  et~al.}(2015{\natexlab{a}})}]{Palanque-Delabrouille:2014jca}
\bibinfo{author}{\bibfnamefont{N.}~\bibnamefont{Palanque-Delabrouille}}
  \bibnamefont{et~al.}, \bibinfo{journal}{JCAP}
  \textbf{\bibinfo{volume}{1502}}, \bibinfo{pages}{045}
  (\bibinfo{year}{2015}{\natexlab{a}}), \eprint{1410.7244}.

\bibitem[{\citenamefont{Palanque-Delabrouille
  et~al.}(2015{\natexlab{b}})}]{Palanque-Delabrouille:2015pga}
\bibinfo{author}{\bibfnamefont{N.}~\bibnamefont{Palanque-Delabrouille}}
  \bibnamefont{et~al.}, \bibinfo{journal}{JCAP}
  \textbf{\bibinfo{volume}{1511}}, \bibinfo{pages}{011}
  (\bibinfo{year}{2015}{\natexlab{b}}), \eprint{1506.05976}.

\bibitem[{\citenamefont{Baur et~al.}(2017)\citenamefont{Baur,
  Palanque-Delabrouille, Yeche, Boyarsky, Ruchayskiy, Armengaud, and
  Lesgourgues}}]{Baur:2017stq}
\bibinfo{author}{\bibfnamefont{J.}~\bibnamefont{Baur}},
  \bibinfo{author}{\bibfnamefont{N.}~\bibnamefont{Palanque-Delabrouille}},
  \bibinfo{author}{\bibfnamefont{C.}~\bibnamefont{Yeche}},
  \bibinfo{author}{\bibfnamefont{A.}~\bibnamefont{Boyarsky}},
  \bibinfo{author}{\bibfnamefont{O.}~\bibnamefont{Ruchayskiy}},
  \bibinfo{author}{\bibfnamefont{Ã.}~\bibnamefont{Armengaud}},
  \bibnamefont{and}
  \bibinfo{author}{\bibfnamefont{J.}~\bibnamefont{Lesgourgues}},
  \bibinfo{journal}{JCAP} \textbf{\bibinfo{volume}{1712}}, \bibinfo{pages}{013}
  (\bibinfo{year}{2017}), \eprint{1706.03118}.

\bibitem[{\citenamefont{Aghanim et~al.}(2018)}]{Aghanim:2018eyx}
\bibinfo{author}{\bibfnamefont{N.}~\bibnamefont{Aghanim}} \bibnamefont{et~al.}
  (\bibinfo{collaboration}{Planck}) (\bibinfo{year}{2018}),
  \eprint{1807.06209}.

\bibitem[{\citenamefont{K{\"o}nig et~al.}(2016)\citenamefont{K{\"o}nig, Merle,
  and Totzauer}}]{Konig:2016dzg}
\bibinfo{author}{\bibfnamefont{J.}~\bibnamefont{K{\"o}nig}},
  \bibinfo{author}{\bibfnamefont{A.}~\bibnamefont{Merle}}, \bibnamefont{and}
  \bibinfo{author}{\bibfnamefont{M.}~\bibnamefont{Totzauer}},
  \bibinfo{journal}{JCAP} \textbf{\bibinfo{volume}{1611}}, \bibinfo{pages}{038}
  (\bibinfo{year}{2016}), \eprint{1609.01289}.

\bibitem[{\citenamefont{Boulebnane et~al.}(2018)\citenamefont{Boulebnane,
  Heeck, Nguyen, and Teresi}}]{Boulebnane:2017fxw}
\bibinfo{author}{\bibfnamefont{S.}~\bibnamefont{Boulebnane}},
  \bibinfo{author}{\bibfnamefont{J.}~\bibnamefont{Heeck}},
  \bibinfo{author}{\bibfnamefont{A.}~\bibnamefont{Nguyen}}, \bibnamefont{and}
  \bibinfo{author}{\bibfnamefont{D.}~\bibnamefont{Teresi}},
  \bibinfo{journal}{JCAP} \textbf{\bibinfo{volume}{1804}}, \bibinfo{pages}{006}
  (\bibinfo{year}{2018}), \eprint{1709.07283}.

\bibitem[{\citenamefont{Garny et~al.}(2018{\natexlab{b}})\citenamefont{Garny,
  Konstandin, Sagunski, and Tulin}}]{Garny:2018byk}
\bibinfo{author}{\bibfnamefont{M.}~\bibnamefont{Garny}},
  \bibinfo{author}{\bibfnamefont{T.}~\bibnamefont{Konstandin}},
  \bibinfo{author}{\bibfnamefont{L.}~\bibnamefont{Sagunski}}, \bibnamefont{and}
  \bibinfo{author}{\bibfnamefont{S.}~\bibnamefont{Tulin}},
  \bibinfo{journal}{JCAP} \textbf{\bibinfo{volume}{1809}}, \bibinfo{pages}{011}
  (\bibinfo{year}{2018}{\natexlab{b}}), \eprint{1805.12203}.

\bibitem[{\citenamefont{Murgia et~al.}(2018)\citenamefont{Murgia,
  Ir\v{s}i\v{c}, and Viel}}]{Murgia:2018now}
\bibinfo{author}{\bibfnamefont{R.}~\bibnamefont{Murgia}},
  \bibinfo{author}{\bibfnamefont{V.}~\bibnamefont{Ir\v{s}i\v{c}}},
  \bibnamefont{and} \bibinfo{author}{\bibfnamefont{M.}~\bibnamefont{Viel}}
  (\bibinfo{year}{2018}), \eprint{1806.08371}.

\bibitem[{\citenamefont{Sigurdson and Kamionkowski}(2004)}]{Sigurdson:2003vy}
\bibinfo{author}{\bibfnamefont{K.}~\bibnamefont{Sigurdson}} \bibnamefont{and}
  \bibinfo{author}{\bibfnamefont{M.}~\bibnamefont{Kamionkowski}},
  \bibinfo{journal}{Phys. Rev. Lett.} \textbf{\bibinfo{volume}{92}},
  \bibinfo{pages}{171302} (\bibinfo{year}{2004}), \eprint{astro-ph/0311486}.

\end{thebibliography}

\end{document}